\documentclass[pra,twocolumn,showpacs,superscriptaddress,amsmath,amssymb]{revtex4}
\usepackage{graphicx}
\usepackage{amssymb}
\usepackage{bm}
\usepackage{braket}
\usepackage{amsmath}

\begin {document}

\title{Geometric phase of a central spin coupled to an
antiferromagnetic environment}
\author{Xiao-Zhong Yuan}
\affiliation{Department of Physics, Shanghai Jiao Tong University,
Shanghai 200240, China}
\author{Hsi-Sheng Goan}\email{goan@phys.ntu.edu.tw}
\affiliation{Department of Physics, and Center for Theoretical
Sciences, National Taiwan University, Taipei 10617, Taiwan} 
\affiliation{Center for Quantum
Science and Engineering, National Taiwan University, Taipei 10617, Taiwan}
\author{Ka-Di Zhu}
\affiliation{Department of Physics, Shanghai Jiao Tong University,
Shanghai 200240, China}

\begin{abstract} Using the spin-wave approximation, we study the
geometric phase (GP) of a central spin (signal qubit) coupled to an 
antiferromagnetic (AF) environment under the application of 
an external global magnetic field. 
The external magnetic field affects the GP of the
qubit directly and also indirectly
through its effect on the AF environment.
We find that when the applied magnetic field is increased to the critical
magnetic field point, the AF environment undergoes a spin-flop
transition, a first-order phase transition, and 
at the same time the GP of the qubit
changes abruptly to zero.
This sensitive change of the GP of a signal
qubit to the
parameter change of a many-body environment near its critical point
may serve as another efficient tool or witness to study the many-body
phase transition. 
The influences of the AF environment 
temperature and crystal anisotropy field on
the GP are also investigated.
\end{abstract}

\pacs{03.65.Vf, 03.65.Yz, 75.30.Ds}
\maketitle

The notion of adiabatic geometric phase (GP) of a quantum system 
first discussed in the
pioneer work of Berry \cite{1}
was generalized and extended to
nonadiabatic \cite{Aharonov87} and noncyclic \cite{Samuel88} evolutions.  
It was also generalized 
to mixed states and nonunitary 
evolutions \cite{10,14,Yi}.
Recently, the close relation between the GP's and 
quantum phase transition (QPT) in many-body systems 
was suggested and investigated \cite{Carollo}.
In the QPT witness studies,
an auxiliary spin or signal qubit is introduced to couple with a many-body 
system (environment) and the notion
of Loschmidt echo (or equivalently the
decoherence factor of the qubit) \cite{Quan06}
or quantum state fidelity \cite{Zanardi06}  
is used to investigate the quantum
criticality. 
The signature of QPT 
is the dramatic decay of the asymptotic value of the Loschmidt echo 
or the quantum state fidelity at the critical
point. That is, the closer the
environment to the QPT, the smaller the
asymptotic value of the Loschmidt echo or the
decoherence factor of the signal qubit. 
Recently, the GP of an auxiliary qubit 
induced by a
one-dimensional $XY$ spin chain (an external environment) was calculated
and the result was used to study the criticality of a 
special case of the transverse Ising model \cite{Yi07}.    
It was found that the induced GP in the qubit 
changes dramatically at the critical point \cite{Yi07}.

Antiferromagnets subjected to an
external magnetic field attracted considerable attention over
the years \cite{Tanaka}.
One of the interesting
phenomena in antiferromagnetic (AF) materials under an applied magnetic
field is the magnetic-field-induced spin-flop transition. When the
applied magnetic field is increased to the critical field point,
the AF polarization flips into the direction
perpendicular to the field. This is called the spin-flop
transition, a first-order phase transition. 
The phenomena of the spin-flop transition were
observed experimentally \cite{21,22}. It is thus particularly
interesting to investigate how a globally applied external field
influences the GP of a signal qubit that is coupled to an 
AF environment 
especially when the magnetic field
strength is near the critical field of the spin-flop transition.
It is known that the values of the
critical magnetic field of the spin-flop transition can be
obtained using the spin-wave theory. We thus apply
the spin-wave approximation to deal with the AF
environment in a globally applied magnetic field
at low temperatures and low-energy excitations.
The spin-wave approach allows us to go beyond the Markovian
approximation and the weak-coupling limit in the usual
treatments for system-coupled-to-environment models. 
The influence of the temperature and the crystal
anisotropy field on the GP will also 
be investigated.

The total Hamiltonian of our model \cite{24,25} can be written as
$H=H_S+H_{SB}+H_B$,
where  
\begin{eqnarray}
H_S&=&-g\mu_BBS_0^z,\\
H_{SB}&=&-\frac{J_0}{\sqrt{N}}S_0^z \sum_{i}
(S_{a,i}^z+S_{b,i}^z),\\
H_B&=&J\sum_{i,\vec{\delta}}\mathbf{S}_{a,i}\cdot\mathbf{S}_{b,i+\vec{\delta}}
+J\sum_{j,\vec{\delta}}\mathbf{S}_{b,j}\cdot\mathbf{S}_{a,j+\vec{\delta}}\nonumber\\
&&-g\mu_B(B+B_A)\sum_{i} S_{a,i}^z-g\mu_B(B-B_A)\sum_{j}
S_{b,j}^z,\nonumber\\
&& 
\end{eqnarray}
are the Hamiltonians of the signal qubit (central spin),
the coupling, and the environment,
respectively.
Her $g$ is the gyromagnetic factor and $\mu_B$ is the Bohr
magneton. For simplicity, a significant interaction between the
central spin and the environment is assumed to be of the Ising
type with $J_0$ being the coupling constant. $J$ is the exchange
interaction and is positive for the AF environment. 
We assume that the spin structure of the environment may be divided
into two interpenetrating sublattices $a$ and $b$ with the
property that all nearest neighbors of an atom on $a$ lie on $b$
and \textit{vice versa}. Each sublattice contains $N$ atoms with spin
$S$, and $\mathbf{S}_{a,i}$ ($\mathbf{S}_{b,j}$)
represents the spin operator of the $i$th ($j$th) atom on
sublattice $a$ ($b$).  The
indices $i$ and $j$ label the $N$ atoms, whereas the vectors
$\vec{\delta}$ connect atom $i$ or $j$ with its nearest neighbors.
$B$ represents a uniform external magnetic field applied in the
$z$ direction. The anisotropy field $B_A$ is assumed to be
positive, which approximates the effect of the crystal anisotropy
energy, with the property of tending for positive magnetic moment
$\mu_B$ to align the spins on sublattice $a$ in the positive $z$
direction and the spins on sublattice $b$ in the negative $z$
direction.

Using the Holstein-Primakoff transformation to map the spin operators of
the AF environment onto bosonic operators,
considering the situation that the environment is
in the low-temperature and low-excitation limits,
then transforming the resultant Hamiltonians to the momentum
space, and finally using the Bogoliubov transformation,
we obtain in the spin-wave approximation \cite{25}
\begin{eqnarray}
H_{SB}&=&-\frac{J_0}{\sqrt{N}}S_0^z \sum_{\mathbf{k}}
(\beta_\mathbf{k}^\dagger\beta_\mathbf{k}-\alpha_\mathbf{k}^\dagger\alpha_\mathbf{k}),\\
H_B&=&\sum_{\mathbf{k}}\omega_\mathbf{k}^{(+)}\left(\alpha_\mathbf{k}^\dagger\alpha_\mathbf{k}+\frac{1}{2}\right)+\sum_{\mathbf{k}}\omega_\mathbf{k}^{(-)}\left(\beta_\mathbf{k}^\dagger\beta_\mathbf{k}+\frac{1}{2}\right),
\label{HBmagnon}
\end{eqnarray}
where $\alpha_\mathbf{k}^\dagger$ ($\alpha_\mathbf{k}$) and
$\beta_\mathbf{k}^\dagger$ ($\beta_\mathbf{k}$) are the creation
(annihilation) operators of the two different magnons with
wave vector ${\mathbf{k}}$ and frequency
$\omega_{\mathbf{k}}^{(+)}$ ($\omega_{\mathbf{k}}^{(-)}$),
respectively, and 
 $\hbar=1$ 
in Eq.~(\ref{HBmagnon}). 
For a cubic crystal system in the small $k$
approximation
\begin{eqnarray}
\omega_\mathbf{k}^{(\pm)}&=&2MSJ\sqrt{\left(1+\frac{g\mu_BB_A}{2MSJ}\right)^2+2\frac{k^2l^2}{M}-1}\pm
g\mu_BB,\nonumber\\
&& 
\end{eqnarray}
where $M$ is the number of nearest neighbors of an atom and $l$ is the side length of cubic primitive cell of the
sublattice. 

We assume that the initial density matrix of the total systems is
separable (i.e., $\rho(0)=|\psi\rangle\langle\psi|\otimes\rho_B$,
where the density matrix of the environment 
$\rho_B=e^{-H_B/T}/Z$, where $Z$ is the partition function and
the Boltzmann constant has been set to one. 
The initial state of the central spin is described by
$|\psi\rangle=\sin(\theta_0/2)|e\rangle+\cos(\theta_0/2)|g\rangle$,
where $|e\rangle$, $|g\rangle$ denote the excited and ground
states of the qubit and $\theta_0$ is the polar angle of the initial
state in the Bloch sphere representation with $|g\rangle$ at the north 
pole.
The reduced density matrix of the qubit in the thermodynamics
limit (i.e., $N\rightarrow \infty$)
 can be obtained, following the
calculation in Ref.~\cite{25}, to be
\begin{eqnarray}
\rho(t)&=&\left(\begin{array}{cc}
\sin^2\frac{\theta_0}{2} & 
\frac{1}{2}\sin\theta_0 \, e^{-ig\mu_BBt-(t/\tau_0)^2} \\
\frac{1}{2}\sin\theta_0\,  e^{ig\mu_BBt-(t/\tau_0)^2} &
\cos^2\frac{\theta_0}{2}
\end{array}\right).\nonumber\\
&& 
\label{rhot}
\end{eqnarray}
The decoherence time $\tau_0$ is given by \cite{25}
$\tau_0={\sqrt{2}\pi}/[{J_0\sqrt{\left(\eta^++\eta^-\right)}}]$,
where
\begin{eqnarray}
\eta^{\pm}&=&\frac{1}{2}\int_0^\infty
\frac{e^{-\omega_\mathbf{k}^{(\pm)}/T}}{(1-e^{-\omega_\mathbf{k}^{(\pm)}/T})^2}x^2dx,
\label{eta_limit}
\end{eqnarray}
with $x=kl$.
The decoherence time depends on the coupling strength to the
environment, on the structure parameters and temperature 
of the environment, and on the external magnetic field. 

To calculate the 
GP for the qubit undergoing nonunitary
evolution, we use the gauge invariant expression derived in \cite{10}
\begin{eqnarray}
\Phi=\arg\left(\sum_{k}\sqrt{\varepsilon_k(0)\varepsilon_k(\tau)}\langle\phi_k(0)|\phi_k(\tau)\rangle
e^{-\int_0^\tau dt\langle\phi_k|\partial/\partial
t|\phi_k\rangle}\right),
\nonumber\\
&& 
\label{geometric_phase}
\end{eqnarray}
where $\tau$ denotes the total evolution time, and 
$\varepsilon_k(\tau)$ and $|\phi_k(\tau)\rangle$ are
the eigenvalues and corresponding eigenvectors of the reduced density
matrix $\rho(t)$. One may view the GP factor defined in 
Eq.~(\ref{geometric_phase}) as a weighted sum over
the phase factors pertaining to the eigenvectors of the reduced 
density matrix. 
The eigenvalues and eigenstates of $\rho(t)$ in Eq.~(\ref{rhot}) can be
obtained as
\begin{eqnarray}
\varepsilon_{\pm}(t)&=&\frac{1}{2}\pm\frac{1}{2}\sqrt{\cos^2\theta_0+e^{-2t^2/\tau_0^2}\sin^2\theta_0}\,
,
\label{eigenvalues}
\\
|\phi_\pm(t)\rangle&=&\frac{1}{\sqrt{\sin^2\theta_0 \, e^{-2t^2/\tau_0^2}+4[\varepsilon_{\pm}(t)-\sin^2\frac{\theta_0}{2}]^2}}
\nonumber\\
&&\times \left\{\sin\theta_0 \, e^{-ig\mu_BBt-t^2/\tau_0^2}|e\rangle\right.\nonumber\\
&&\qquad\left.+2[\varepsilon_{\pm}(t)-\sin^2\frac{\theta_0}{2}]|g\rangle\right\}.
\nonumber\\
&& 
\label{eigenvectors}
\end{eqnarray}
Substituting Eqs.~(\ref{eigenvalues}) and (\ref{eigenvectors}) into
Eq.~(\ref{geometric_phase}) and 
carrying out some simple manipulations, we obtain the GP acquired at
the time $\tau=2\pi/(g\mu_B B )$ of a cyclic period
\begin{eqnarray}
\Phi
&=&\int_0^\tau
dt\frac{g\mu_BB\sin^2\theta_0 \, e^{-2t^2/\tau_0^2}}{\sin^2\theta_0 \, e^{-2t^2/\tau_0^2}+4(\varepsilon_{+}-\sin^2\frac{\theta_0}{2})^2}.
\label{eq1}
\end{eqnarray}
The AF environment causes decoherence and
influences the GP of the signal qubit. 
Our results obtained
using the spin-wave approximation are valid only for not too large
temperatures, much smaller than the N\'{e}el temperature.
According to neutron diffraction studies, the N\'{e}el temperature
of antiferromagnet TbAuIn is $35\,\mathrm{K}$ \cite{49}. Some
antiferromagnets may have higher N\'{e}el temperatures.  So in the
following analysis, the environmental temperature is restricted
below $T/g\mu_B=2.5\,\mathrm{Tesla}$ (i.e., $T\approx
3.4\,\mathrm{K}$).

We, next, discuss the GP acquired at 
the time $\tau=2\pi/(g\mu_B B )$ after the qubit completes a
cyclic evolution when it is isolated from the AF
environment. 
If the qubit is really isolated
from the AF environment, i.e., $J_0=0$, then
the decoherence time $\tau_0\to \infty$. 
In this case, we obtain from Eq.~(\ref{eq1}) the well-known accumulated 
 geometric
phase of a spin-${1}/{2}$ particle 
processing around an external magnetic field in each cycle as:
$\Phi=\pi(1-\cos\theta_0)$.
If $\theta_0=\frac{\pi}{2}$, we have $\Phi=\pi$. 

\begin{figure}[t]
\begin{center}
\includegraphics [width=\linewidth] {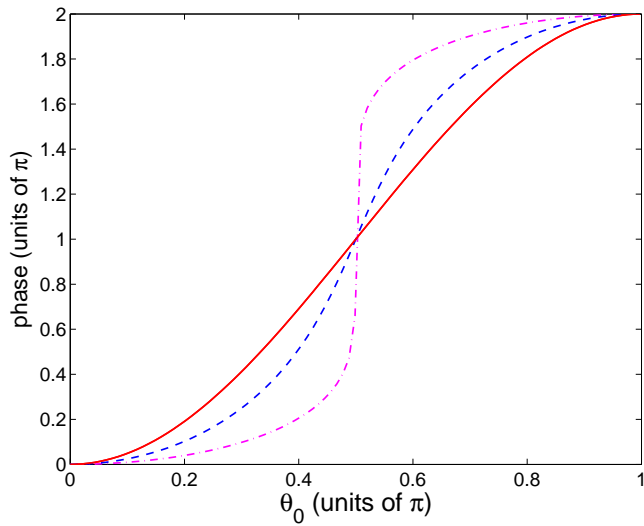}
\end{center}
\caption{(Color online) The GP of the signal qubit as a function of
the polar angle $\theta_0$ for
$J_0=0$ (solid curve), $J_0=2.5J$,
$T/g\mu_B=0.8\,\mathrm{Tesla}$ (dashed curve), and
$J_0=2.5J$, $T/g\mu_B=1.2\,\mathrm{Tesla}$ (dot-dashed curve). Other
parameters are $M=6$, $MJ/g\mu_B=40\,\mathrm{Tesla}$,
$B_A=0.10\,\mathrm{Tesla}$, and $B=0.5\,\mathrm{Tesla}$}.
\label{fig1}
\end{figure}

Figure \ref{fig1} shows the GP of the qubit as a function of 
the initial polar angle $\theta_0$ for
different temperatures.  The external magnetic field is
chosen as $B=0.5\,\mathrm{Tesla}$  
and other parameters are with $B_A=0.1\,\mathrm{Tesla}$  and
$MJ/g\mu_B=40\,\mathrm{Tesla}$.
The solid curve corresponds to
the qubit isolated from the AF environment ($J_0=0$) [i.e.,
$\Phi=\pi(1-\cos\theta_0)$]. 
Comparing with the isolated case, 
we can see from Fig.~\ref{fig1} that
the influence of the environment makes the GP
curve deviate steeply from the isolated case in the
region around $\theta_0={\pi}/{2}$. 
The deviation of the GP is positive for $\theta_0>\pi/2$  
and is negative for $\theta_0<\pi/2$. 
Furthermore, the higher the temperature
is, the steeper the curve is. 
For three special initial states
with polar angles 
$\theta_0=0$, $\theta_0=\frac{\pi}{2}$, and $\theta_0=\pi$, the
GP's are not affected by the  AF environment.

\begin{figure}[b]
\begin{center}
\includegraphics  [width=\linewidth] {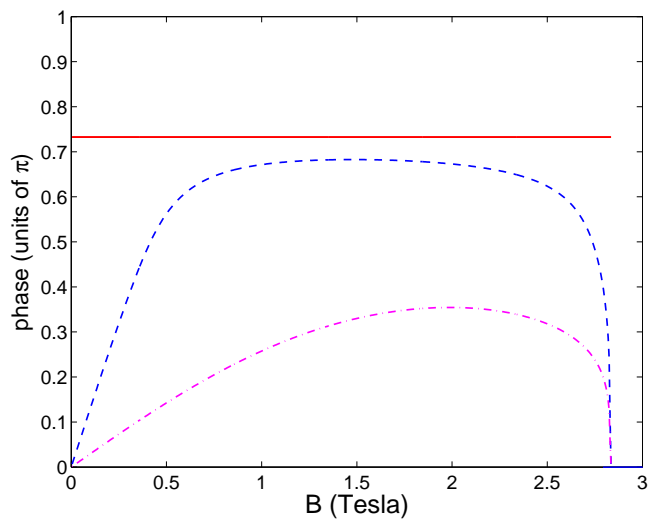}
\end{center}
\caption{(Color online) The GP of the signal qubit as a function of
the external magnetic field $B$ for 
$J_0=0$ (solid curve), $J_0=2.5J$,
$T/g\mu_B=0.8\,\mathrm{Tesla}$ (dashed curve), and
$J_0=2.5J$, $T/g\mu_B=1.5\,\mathrm{Tesla}$ (dot-dashed curve). 
Other parameters are $\theta_0=1.3$, $M=6$,
$MJ/g\mu_B=40\,\mathrm{Tesla}$, and $B_A=0.10\,\mathrm{Tesla}$.}
\label{fig2}
\end{figure}
\begin{figure}[b]
\begin{center}
\includegraphics  [width=\linewidth] {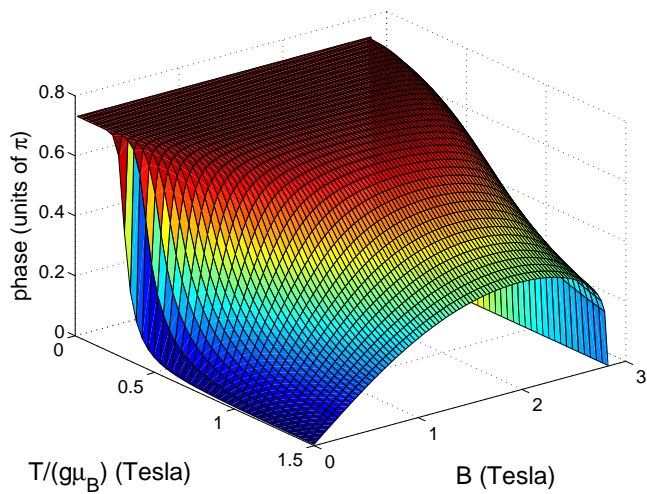}
\end{center}
\caption{(Color online) The GP of the signal qubit versus the
external magnetic field $B$ and the environment temperature $T$. 
Other parameters are $\theta_0=1.3$, $M=6$,
$MJ/g\mu_B=40\,\mathrm{Tesla}$, $J_0=2.5J$, and
$B_A=0.10\,\mathrm{Tesla}$.} \label{fig3}
\end{figure}

In Fig.~\ref{fig2}, we plot the GP as a function of external
magnetic field $B$ for different temperatures with $\theta_0=1.3$.
The solid curve corresponds to the isolated case ($J_0=0$)
[i.e., $\Phi=\pi(1-\cos1.3)=0.73\pi$], which is independent of $B$
for $B\neq 0$. It is obvious from Fig.~\ref{fig2}
that the curve with higher temperature
shows greater deviation from $\Phi=0.73\pi$. 
For a given strength of the 
external magnetic field, the decoherence time $\tau_0$
of the off-diagonal elements of the reduced density matrix decreases with
the increase of temperature. This reduces the GP
obtained by the qubit \cite{Yi}. 
The influence of the external magnetic field
on the GP (\ref{eq1}) 
can be categorized into two competing ways: one through
the  decoherence time $\tau_0$  
and the other one
through the quasiperiod $\tau={2\pi}/({g\mu_BB})$.
If the external
magnetic field increases, 
the decoherence time $\tau_0$ 
decreases \cite{25}. Then it suggests that the influence of the
environment on the GP increases with 
the increasing magnetic field. 
On the other hand, with the increase of the external magnetic
field $B$, the quasiperiod $\tau$ 
decreases. It then takes less time for the
environment to influence the GP in a quasicycle. 
In this sense, the 
influence of environment on the GP decreases with the
increase of the external magnetic field. 
Hence, when the external magnetic field is very small, the 
decoherence time $\tau_0$ is finite  \cite{24,25} but
the qubit needs a rather long time $\tau\gg\tau_0$ 
to complete a quasicycle. 
Since the integrand with the factor $\exp(-2t^2/\tau_0^2)$ in Eq.~(\ref{eq1})
becomes very small for time $t$ considerably larger than $\tau_0$, 
the integration over $t$ from $\tau_0$ to $\tau$
thus contributes very little to the whole GP of the signal qubit
as compared to the 
isolated case.
As a result, the
dashed and dot-dashed curves at very
small $B$ 
show large deviation from the value $\Phi=0.73\pi$ 
of the isolated case.  
One can furthermore obtain from  Eq.~(\ref{eq1}) that 
the GP near $B=0$ 
behaves linearly in $B$ as
$\lim\limits_{B \rightarrow 0}\Phi=c_1 B$,
where $c_1$ is independent of $B$.
As one increases the strength of the magnetic filed,
the decoherence time decreases but the quasiperiod also decrease.
Usually, the
second effect is more important in typical values of the
external magnetic field.  
Therefore, the low-temperature dashed curve 
approaches $\Phi=0.73\pi$ with the increase of the external
magnetic field.
When the strength of the magnetic field is increased
further toward the critical point where the spin-flop transition
occurs, $B=B_c\approx2.83\,\mathrm{Tesla}$ for the parameters used
in Figs.~\ref{fig2} and \ref{fig3}, 
the decoherence time approaches zero abruptly \cite{25}. 
An analytical expression for the decoherence time near the
critical field can be found in Eq.~(39) of Ref.~\cite{25}. 
We note that this result using spin-wave theory is valid for
$B<B_c$ (i.e., before the spin-flop transition).
Thus when $B\to B_c$ from below, 
the environment exerts a great influence on the qubit
so that its coherence is destroyed completely,
but the quasiperiod is still finite.
Such a great influence thus
also appears on the GP and causes a sudden deviation
from $\Phi=0.73\pi$ for the dashed curves in Fig.~\ref{fig2} when the
external magnetic field approaches $B_c=2.83$ Tesla.
The drastic change in the induced GP of the qubit at the
critical point is due to the sensitivity of the many-body (AF)
environment to the parameter (the magnetic field) change 
near its critical point.   
Furthermore, 
from Eq.~(39) of Ref.~\cite{25} and Eq.~(\ref{eq1}), 
one can
obtain the GP near the critical field scaling as
$\lim\limits_{B \rightarrow B_c(B<B_c)}\Phi=c_2
(B_c-B)^{1/4}$.
Here $c_2$ is independent of the external magnetic field $B$.
The linear behaviors of the GP near $B=0$ and
the abrupt changes near $B=B_c$ can be clearly observed in Fig.~\ref{fig2}.

\begin{figure}[t]
\begin{center}
\includegraphics  [width=\linewidth] {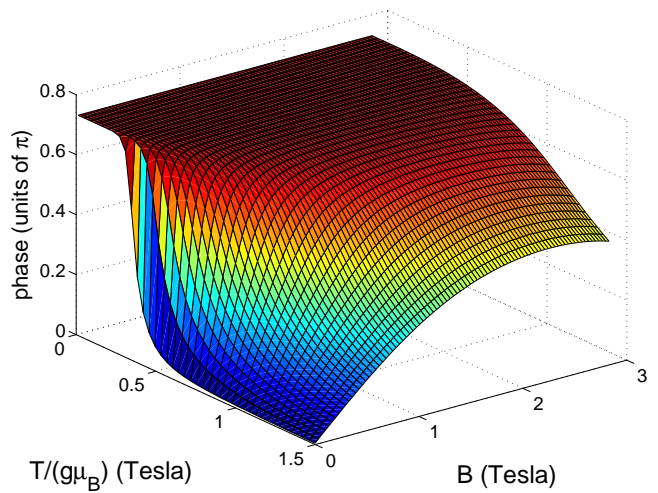}
\end{center}
\caption{(Color online) Same as Fig. 3, except $B_A=0.15\,\mathrm{Tesla}$.}
\label{fig4}
\end{figure}

In Fig.~\ref{fig3}, the GP of the qubit 
(central spin) is plotted as
a function of $B$ and $T$ for the value of anisotropic field
$B_A=0.10$ $\mathrm{Tesla}$. 
Obviously, the curves in Fig.~\ref{fig2} are cross-section curves of the
curved surface in Fig.~\ref{fig3} at different temperatures. As
mentioned previously, one can observe from Fig.~\ref{fig3} that
the GP decreases from the value $\Phi=0.73\pi$
with the increase of the temperature. Besides, 
for $T/g\mu_B>0.5\,\mathrm{Tesla}$, 
it decreases quickly when $B$
approaches zero. As the external magnetic field 
approaches the critical point of 
$B_c\approx2.83\,$ $\mathrm{Tesla}$, the GP changes
drastically to zero.
Figure \ref{fig4} is similar to  Fig.~\ref{fig3}, 
except that the
anisotropy field is at $B_A=0.15\, $ $\mathrm{Tesla}$. Then the
spin-flop transition occurs at $B=B_c\approx3.47\, $ 
$\mathrm{Tesla}$. In such a case, the GP does not show
abrupt change even when $B$ approaches 3 $\mathrm{Tesla}$.
Comparing Fig.~\ref{fig3} with Fig.~\ref{fig4}, 
we find that the large crystal
anisotropy field suppresses the influence of the AF 
environment on the GP.

In conclusion, we investigated the influence of an
AF environment on the GP of a signal qubit
(central spin). 
Such an influence is enhanced by increasing the temperature or 
decreasing the crystal anisotropy field. 
The dependence of the GP on the external magnetic field
$B$ involves two competing factors: quasiperiod $\tau$ and
decoherence time $\tau_0$. 
A larger $B$ implies
a smaller $\tau$ which indicates that the GP acquired at time
$\tau$ by the qubit is less influenced by the environment in terms of
the duration of interaction time with the environment.
But the larger $B$ also implies smaller $\tau_0$ which
indicates that the decays in the integrand of the 
GP (\ref{eq1}) become greater. 
When $B$ approaches the region near the critical field of the spin-flop
transition,  $\tau_0\to 0$ but $\tau$ is finite. 
The qubit acquires no GP with $\tau_0 \to 0$ and thus
the GP
changes abruptly at the critical field of the spin-flop
transition. This sensitive change of the GP of a signal
qubit to the
parameter change of a many-body environment near its critical point
may serve as, in addition to the Loschmidt echo
and quantum state fidelity, 
another efficient tool or witness to study the phase transition.

H.S.G. acknowledges support from 
NSC under Grant No.~97-2112-M-002-012-MY3, 
from NTU 
under Grants Nos.~97R0066-65 and 
97R0066-67,
and from the NCTS focus group program. 
X.Z.Y. acknowledges support
from NNSF of China under Grant
No. 10874117. 
X.Z.Y. also thanks the conference support of ICTP (smr2046).



\end{document}